\documentclass{mem}
\usepackage{natbib}
\usepackage{txfonts}
\usepackage{balance}
\usepackage{graphicx}
\idline{00}{000}

\begin{document}
\def\teff{$T\rm_{eff }$}
\def\kms{$\mathrm {km s}^{-1}$}

\title{Improved implementation of dust-driven winds and dust formation in models of AGB evolution:}
   \subtitle{Effects of pulsation and gas-pressure forcing}
\author{Lars \,Mattsson$^1$ \& Paolo Ventura$^2$}

\institute{
$^1$Nordita, KTH Royal Institute of Technology and Stockholm University, Roslagstullsbacken 23, SE-106 91, Stockholm, Sweden\\
$^2$INAF- Osservatorio Astronomico di Roma, via di Frascati 33, I-00040 Monteporzio, Italy\\
\email{lars.mattsson@nordita.org}
}
\authorrunning{Mattsson}
\titlerunning{Improved win3ds and dust formation in models of AGB evolution}

\abstract{Mass loss is a crucial component in stellar evolution models, since it largely determines the rate of evolution at the later stages of a star's life. The dust-driven outflows from AGB stars are particularly important in this regard. Including AGB dust formation in a stellar evolution model does also require a model of these outflows. Since AGB stars exhibit large-amplitude pulsation, a model based on time-dependent radiation hydrodynamics (RHD) is needed in order to capture all the important physical aspects of dust formation. However, this cannot be afforded in a stellar evolution model. Here, a mean-flow model is presented, which include corrections to the steady-state model currently being used in AGB evolution models with dust formation.

}
\maketitle{}

\section{Introduction}
Radiation pressure on dust grains is usually considered the main driving mechanism behind the winds of asymptotic giant branch (AGB) stars \cite{Jura86,Sedlmayr95,Habing03}. A first attempt to include stellar dust production in stellar evolution modelling was made by Ferrarotti \& Gail (2006) \nocite{Ferrarotti06}, where a synthetic model of AGB evolution was combined with a model of dust formation and a prescribed stationary mass loss as in Gail \& Sedlmayr (1987; 1988), but without thermal gas pressure. Ventura et al. (2012a; 2012b)\nocite{Ventura12a,Ventura12b} used this approach solving the full set of stellar evolution equations, finding that previous theoretical dust yields of AGB stars were overestimated. The contribution of AGB stars to the cosmic dust budget is, in fact, much less than that of interstellar dust condensation together with supernova produced dust \cite[despite significant uncertainties][]{Mattsson15b}, which can be seen in the local Universe as well as at high redshifts \cite[e.g.][]{Mattsson14,Valiante17}. This picture is also  supported by observed dust-depletion patterns \cite{Jenkins09,DeCia16}. 

Neglecting the effects of pulsation and gas pressure is justifiable only as long as the ratio of radiative to gravitational acceleration ${\Gamma}\gg 1$. Moreover, on the one hand, a prescribed mass-loss rate enables dust formation being calculated as ``post processing'', but on the other hand, there is no effect from dust formation on the mass-loss rate. This can be dangerous if the prescription is not reliable and, unfortunately, there is no truly reliable and general prescription. An attempt to improve upon this situation for carbon-rich AGB stars was made by Mattsson et al. (2010) \nocite{Mattsson10}, demonstrating the importance of treating the mass-loss rate as a function of the abundance of free carbon. Grain sizes has also turned out to be an important parameter \cite{Mattsson11}, which further complicates the picture. 

Combining detailed wind modelling with stellar evolution modelling is computationally not feasible, however. Therefore, it is of interest to find an acceptable way to actually {\it model} the mass-loss rate, but at a low computational cost. In this paper, a mean-flow model of a dust-driven stellar wind is discussed, in which the net effect of the pulsation is taken into account in terms of an additional ``wave pressure''. 

\section{Mean-flow equations}
Using Favr\'e (1962) averaging the equation of motion (EOM) for a spherical wind will receive an additional ``wave pressure'' term  (see Mattsson 2016 and references therein for further details),
\begin{equation}
\label{eom2}
\nonumber
\tilde{u}_r {d \tilde{u}_r\over d r}  = -{1\over \bar{\rho}}{d\bar{P}\over dr}  - {GM_\star\over r^2}\left(1-\widetilde{\Gamma}\right) + f_{\rm wp},
\end{equation}
\begin{equation}
 f_{\rm wp} = - {1\over r^2 \bar{\rho}}{d\over dr}\left[r^2 \overline{\rho\,(u_r^{\prime\prime})^2}\right],
\end{equation}
where $\bar{\rho}$ and $\bar{P}$ are the straight time averages of the gas-mass density $\rho$ and thermal pressure $P$, repectively, while $\tilde{u}_r$ is the $\rho$-weighted average of the radial velocity $u_r$, such that $\bar{\rho}\,\tilde{u}_r = \overline{\rho\,u_r} =\overline{\rho\,(\tilde{u}_r + u_r^{\prime\prime})}$, with $u_r^{\prime\prime}$ the time-dependent component of $u_r$. $\widetilde{\Gamma}$ is the corresponding weighted average of $\Gamma$. Mass conservation will lead to $r^2\,\bar{\rho}\,\tilde{u}_r = $~constant, analogous to the exactly stationary case.

The difficult part of this model is to find an expression for $f_{\rm wp}$ which leads to a closed system of equations for the mass-loss problem. Such a relation can be obtained from the energy equation, but will also require some calibration against detailed numerical wind models (this goes beyond the scope of this comprehensive summary, though). However, it is easily shown that $f_{\rm wp}\geq 0$, which means that the EOM has two more terms which increases the outflow velocity compared to the EOM adopted in previous work \cite[see, e.g.,][]{Ferrarotti06,Ventura12a,Ventura12b,Nanni13}.

            \begin{figure*}
  \resizebox{\hsize}{!}{
   \includegraphics{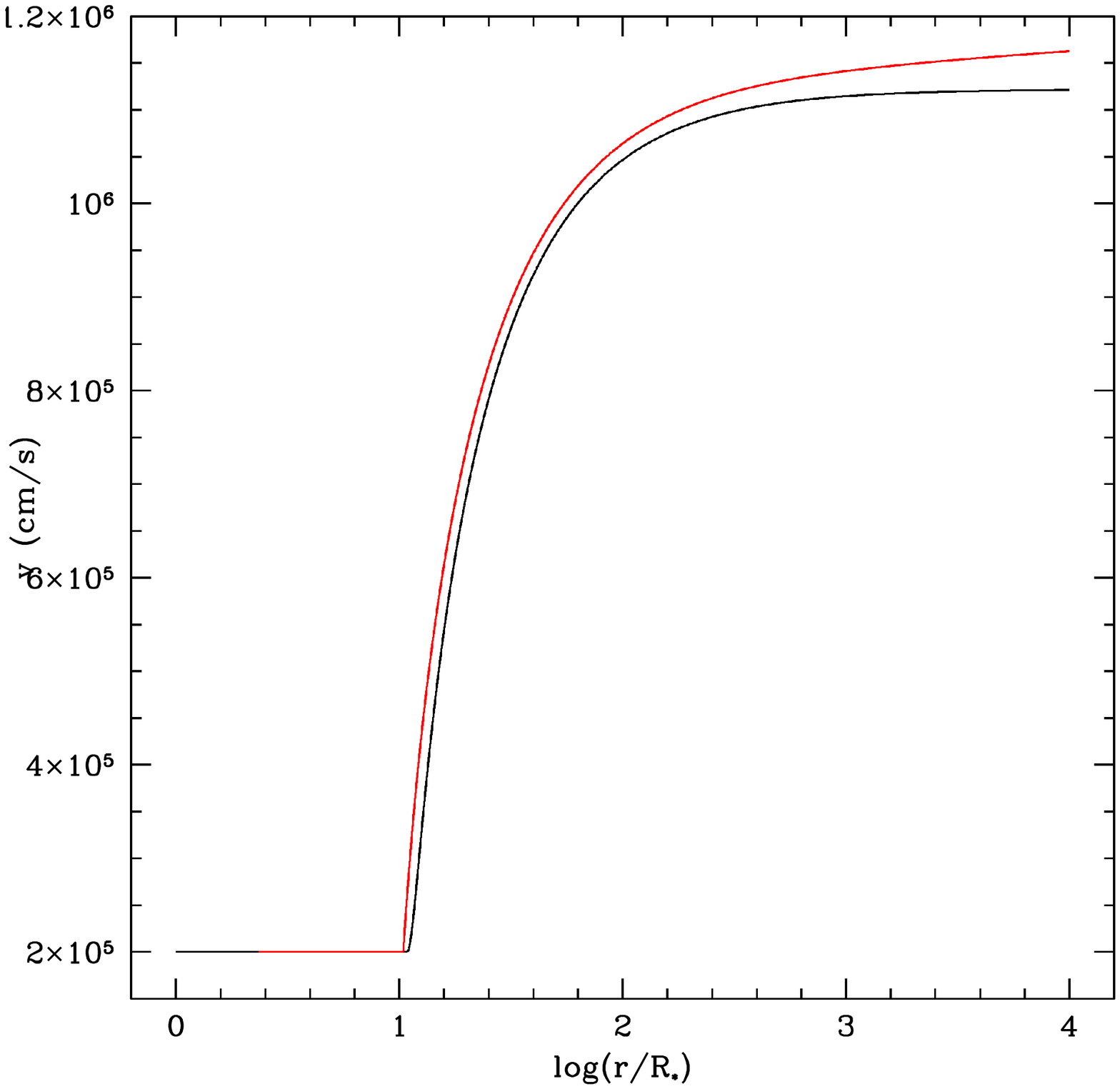}
   \includegraphics{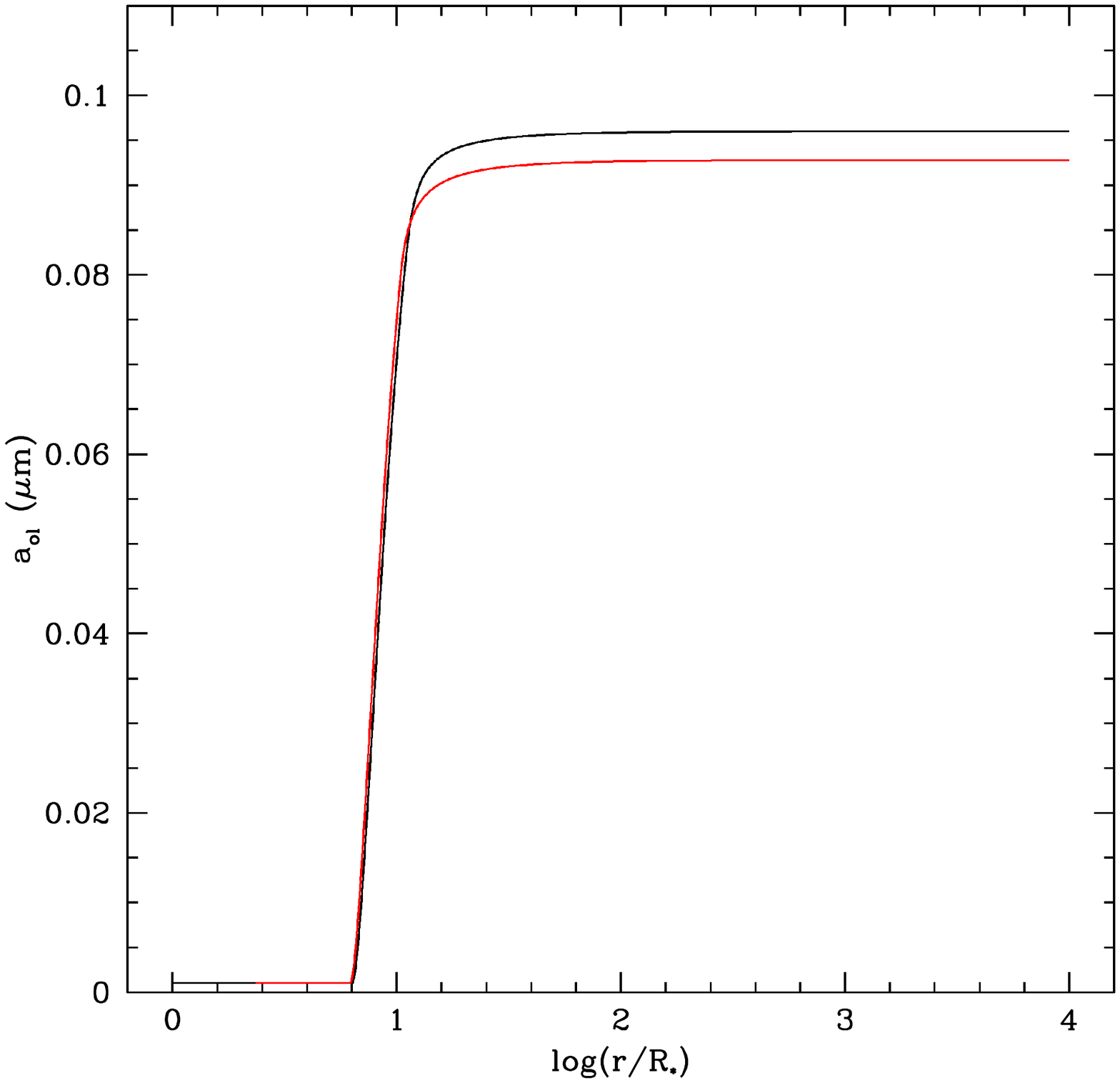}}
  \caption{\label{grainsize} Left: wind-velocity profiles (note: cm~s$^{-1}$) with (red) gas pressure and without (black). Inside the condensation radius the wind velocity is set to 2~km~s$^{-1}$ to avoid numerical issues. Right: increase of the grain radius as a function of distance from the centre of the star. The examples shows the case of olivine-type dust.}
  \end{figure*}

\section{Effects on dust formation}
If one assumes the mass-loss rate $\dot{M}$ in previous calculations of AGB evolution with dust were correctly prescribed, there will be less need for radiation pressure on dust to sustain a wind when the two new pressure terms are added, i.e., $\widetilde{\Gamma} < \Gamma$. In the small-particle limit $\Gamma \propto \rho_{\rm d}/\rho$, where $\rho_{\rm d}$ is the mass density of dust in the wind, which is suggesting the amount of dust needed to reach a given $\dot{M}$ is less.

Fig. \ref{grainsize} shows the increase of the mean grain radius with distance from the centre of the star for olivine dust at some arbitrary time step during the hot-bottom burning phase of a model with initial mass $M_\star=5\,M_{\sun}$, using an updated version of the model by Ventura et al. (2012a; 2012b). The black line is the result without any pressure terms, while the red line shows the mean grain radius with gas pressure included. Both cases assume the same $\dot{M}$. Indeed, the introduction of a pressure gradient generates a non-negligible force which lessens the need for dust formation to sustain an outflow. The grain radius at the considered time step is reduced by a moderate 4\% when gas pressure is added. If this difference in grain radius is characteristic for the whole evolution, then the olivine yield for this model star is 12\% lower than the yield computed without the gas-pressure term. 

The result discussed above is qualitatively robust and is likely to be amplified if pulsation-wave pressure is introduced. It remains unclear to what extent this will directly affect the acceleration of the gas, though.  Detailed modelling has shown that the kinetic-energy input by pulsation is important for the development of a wind \cite{Mattsson08,Mattsson15}, but it is difficult to assess what is direct acceleration due to wave pressure and what is the effect of enhanced conditions for dust formation due to gas compression and ``lifting'' of the atmosphere. 

However, it can be shown that both thermal gas pressure and pulsation-wave pressure will affect the temperature profile of the wind region, which in turn has bearing on the efficiency of dust condensation (assuming the condensation temperature remains roughly the same). In the Lucy approximation, an equation for $T^4$ can be obtained as
\begin{equation}
\label{lucy}
{d\over dr}\left[\left({T\over T_{\rm eff}}\right)^4 \right] = {dW\over dr} - \rho\,\kappa_{\rm tot}\left({R_\star\over r}\right)^2
\end{equation}
where $W$ is the geometric dilution factor ($R_\star$ is the photospheric radius), $\kappa_{\rm tot}$ is the total opacity and $T_{\rm eff}$ is the effective (luminosity) temperature. The radiative-to-gravitional acceleration ratio $\Gamma$ is proportional to $\kappa_{\rm tot}$, i.e.,
\begin{equation}
\kappa_{\rm tot} \approx {4\pi\,c\,GM_\star \Gamma \over L_\star}.
\end{equation}
which combined with Eq. (\ref{lucy}) yields
\begin{equation}
{d\over dr}\left({T^4-\widetilde{T}^4\over T_{\rm eff}^4}\right) = 3\pi\,GM_\star {\bar{\rho}\over L_\star}\left({R_\star\over r}\right)^2 (\Gamma - \widetilde{\Gamma}),
\end{equation}
where $\widetilde{T}$ and $\widetilde{\Gamma}$ are the Favr\'e-averages of $T$ and $\Gamma$ with pressure terms, respectively.
Using $L_\star = 4\pi\,R_\star^2 \sigma_{\rm SB}\, T_{\rm eff}^4$ one then obtains
\begin{equation}
{d\over dr}\left({T^4-\widetilde{T}^4}\right) = {3\over 4}{\bar{\rho}\over \sigma_{\rm SB}}{GM_\star\over r^2}  (\Gamma - \widetilde{\Gamma}),
\end{equation}
which combined with the EOM (assuming same mass-loss rate and mean wind-velocity profile) leads to
\begin{equation}
{d\over dr}\left(T^4-\widetilde{T}^4\right) = {3\over 4\,\sigma_{\rm SB}}
\left({d\bar{P}\over dr}  - \bar{\rho}\,f_{\rm wp} \right).
\end{equation}
The right-hand-side (RHS) of this equation is zero for exactly stationary winds without the pressure terms. For all other (mean-flow) cases it is negative for all $r$.

\section{Modelling of the mass-loss rate}
The mean-flow model of a wind of an AGB star is based on a system of equations which consist of the following:
\begin{itemize}
\item equation of continuity (mass conservation);
\item eqaution of motion (momentum conservation);
\item closure relation for the wave pressure; 
\item radiation/temperature equation (Lucy approximation);
\item moment equations/hierarchy for dust formation.
\end{itemize} 
With appropriate boundary conditions, this system can be solved numerically in order to obtain $\bar{\rho}$ and $\tilde{u}_r$ as functions of $r$. The mass-loss rate is then obtained as $\dot{M} = 4\pi\,r^2\,\bar{\rho}\,\tilde{u}_r$. At the inner boundary, and out to the condensation radius $R_{\rm cond}$ (where dust formation becomes efficient), it is reasonable to assume some kind of quasi equilibrium, i.e., the RHS of Eq. (\ref{eom2}) vanishes, while $\widetilde{\Gamma} = 0$ for an inner-boundary radius $R_{\rm in}< r \le R_{\rm cond}$. In the wind region, i.e., $r> R_{\rm cond}$, the 
RHS of Eq. (\ref{eom2}) should be positive and a regular solution for $\bar{\rho}$ and $\tilde{u}_r$ is therefore obtainable. Thus, the mass-loss rate does not have to be prescribed -- it can be calculated because the EOC provides an equation for $\bar{\rho}$, which can be solved together with the rest of the system, provided that appropriate (inner) boundary conditions are given.

\section{Conclusions}
In this paper it has been argued that:
\begin{enumerate}
\item thermal gas pressure and pulsation are important;
\item less dust will be produced because less dust is needed;
\item mass-loss rate can be modelled (as opposed to prescribed).
\end{enumerate} 
These three points are essential for the calculation of dust yields based on models of  AGB evolution. In particular, the gas pressure, and to some extent also pulsation-wave pressure, must add to the outward acceleration of the atmospheric gas. Hence, the dust yields are expected to become lower than existing dust yields from stellar evolution modelling.

\bibliographystyle{aa}

\end{document}